# Practical Strategies for Integrating a Conversation Analyst in an Iterative Design Process

Allison Woodruff, Margaret H. Szymanski, Rebecca E. Grinter, and Paul M. Aoki
Palo Alto Research Center
3333 Coyote Hill Road
Palo Alto, CA 94304-1314  USA

**ABSTRACT**
We present a case study of an iterative design process that includes a conversation analyst.  We discuss potential benefits of conversation analysis for design, and we describe our strategies for integrating the conversation analyst in the design process.  Since the analyst on our team had no previous exposure to design or engineering, and none of the other members of our team had any experience with conversation analysis, we needed to build a foundation for our interaction.  One of our key strategies was to pair the conversation analyst with a designer in a highly interactive collaboration.  Our tactics have been effective on our project, leading to valuable results that we believe we could not have obtained using another method.  We hope that this paper can serve as a practical guide to those interested in establishing a productive and efficient working relationship between a conversation analyst and the other members of a design team.

**INTRODUCTION**
Social science research methods are often applied in the design of interactive systems.  In particular, qualitative methods such as ethnography can provide useful insights about the environments into which a new system will be introduced.  However, significant difficulties frequently arise in applying rich, highly-contextual qualitative findings to the process of designing and building a concrete artifact.  There are many mismatches between social sciences and other disciplines such as engineering or design – professional vocabulary, accustomed time scale, goals of inquiry, even philosophical outlook [23]; when combined with the time limitations endemic in engineering environments, these mismatches typically (at least in the reported literature – see, e.g., [19]) result in fairly restricted forms of engagement between social scientists and designers.

In the course of a recent project, we were strongly motivated to incorporate a particular qualitative method, *conversation analysis* (CA) [21], into our design process.  The project goal was to build an electronic audio guidebook that would enhance social interaction between tourists.  CA focuses on the processes by which humans organize their activities and interactions; as such, CA has obvious potential to inform the design of computer systems that affect human-human interaction.  While the CA literature has been used as a source of design principles in HCI (e.g., [10]), actually applying CA *as a method* in iterative design seems far more problematic.  CA involves extended, detailed examination of recorded interactions; the analytic framework emphasizes objectively-observed behavioral instances rather than interpretations of behavior by the analyst; and the resulting findings are often quite subtle and require considerable background knowledge to understand.

In this paper, we share our experiences in (successfully) integrating a conversation analyst into an iterative design process.  We believe this report will be useful to others for at least two reasons.  First, we do not know of a previous detailed discussion of this topic.  Second, we describe a process that proceeded "from the ground up" – the conversation analyst on our team had no previous exposure to design or engineering, and none of the other members of our team had any experience with conversation analysis.

We report the methods by which we have integrated the conversation analyst in the design process.  For example, the conversation analyst is paired with the designer in a highly interactive collaboration: the conversation analyst and the designer often work together on both the conversation analysis and the design.

Using the methods to be described, we have found conversation analytic methods to be enormously valuable to our work.  While CA is time- and resource-intensive, we have found ways to manage these costs, and we believe that the analysis has yielded results that we would not have achieved using another method.  We hope that this paper can serve as a practical guide to those interested in establishing a productive and efficient working relationship between a conversation analyst and design team, and in particular, between a conversation analyst and a designer.

The remainder of the paper is organized as follows.  Since this paper is about experiences as well as methods, we provide some necessary background by describing relevant aspects of our project.  We then describe conversation analysis and how design teams might profit from its use, turning to an extended example of how CA has helped improve our own system. In the following section, we discuss the strategies we have used to integrate the conversation analyst in the team to achieve these benefits. Finally, we discuss related work and conclude.

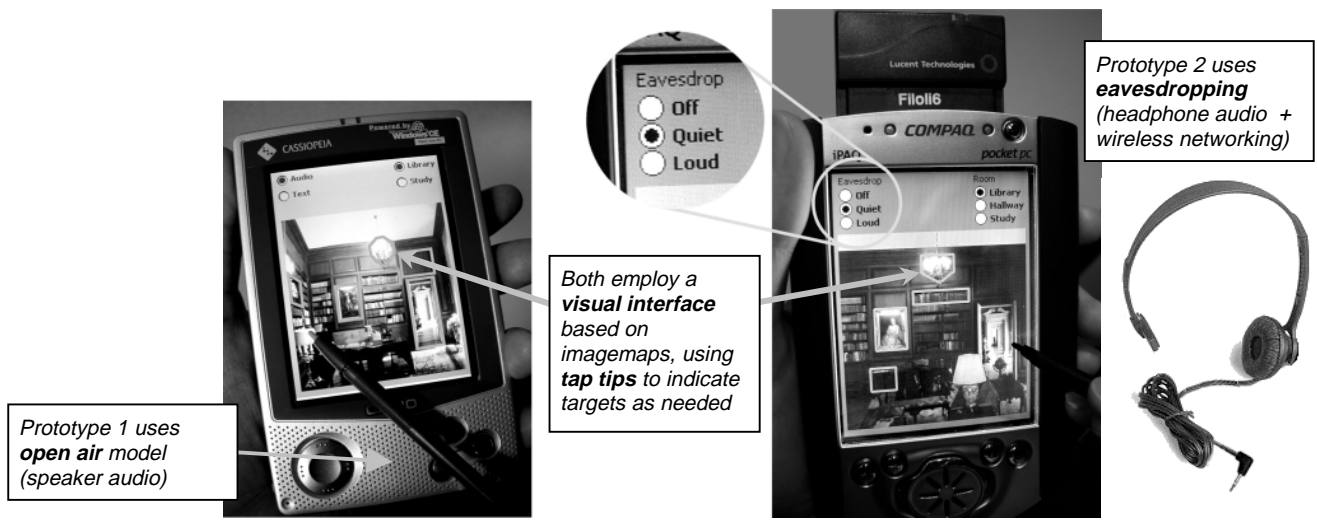

**Figure 1. Comparison of Prototype 1 (left) and Prototype 2 (right).**

**THE ELECTRONIC GUIDEBOOK PROJECT**

To provide context for the rest of the paper, we briefly describe our project, including the team members involved, the prototypes developed, and the studies conducted. Most of this information has been reported previously and in more detail [2,27].

**Team**

Over a two-year period, six members of Xerox PARC have worked intermittently on the electronic guidebook project: an interaction designer, a conversation analyst, a fieldworker, two computer systems researchers, and a computer science intern. None of these members are full-time on this project, and some of these members joined the project after it was already underway; of specific relevance is that the designer was one of the founders of the project, and the conversation analyst joined the project about half a year after it began. The design work is primarily done by the designer (who has an interdisciplinary background, including computer science) in conjunction with one of the computer systems researchers and the conversation analyst, although many design meetings involve all members of the team.

**Prototypes**

We have developed two main prototypes (Figure 1). Each prototype has two main components, a visual interface (which is the same in each prototype) and an information delivery mechanism (which is different in the two prototypes).

*Prototype 1.* At the beginning of the project, we performed a task analysis and developed a design for a visual interface [3]. Individual visitors use this interface to obtain information about objects in their environment. The interface resembles a set of Web browser imagemaps; at a given time, the visitor sees a single photographic imagemap that depicts one wall of a room in a historic house. When visitors tap on an imagemap target, the guidebook delivers a description of that object.

In Prototype 1, visitors had a choice of information delivery modes: text descriptions, audio descriptions played through headphones, or audio descriptions played through speakers. Prototype 1 users predominantly chose audio played through speakers.

*Prototype 2.* The main difference between Prototype 1 and Prototype 2 lies in the information delivery mechanism. Realizing that audio played through speakers is not a feasible solution in a public space due to noise disturbance issues, we moved in Prototype 2 to a single information delivery mode: an audio *eavesdropping* model that uses headsets but preserves the shared listening capability of audio played through speakers. Specifically, devices are paired. Each visitor in a pair always hears the content they select themselves, and additionally, each visitor has a volume control for determining how loudly they hear content from their companion. The volume can be set to "Off," "Quiet," or "Loud" ("Loud" being the same volume as clips selected on one's own device). A priority model addresses overlapping clips; if visitors have selected a clip themselves, they can always hear it. When they are not listening to a clip themselves, they hear other content from their companion's guidebook if (1) they have the volume control set to listen to their companion and (2) their companion is playing a clip.

**Studies**

We believe social interaction is best studied in naturalistic situations. Accordingly, we have conducted three major studies of actual historic house visits. All three studies took place at Filoli, a Georgian revival historic house in Woodside, California (http://www.filoli.org/). In all studies,

| Study 1 | Prototype 1 (speaker audio) |
| --- | --- |
| | 14 visitors |
| | Private visit |
| Study 2 | Prototype 2 (eavesdropped audio) |
| | 12 visitors |
| | Private visit |
| Study 3 | Prototype 2 (eavesdropped audio) |
| | 47 visitors |
| | Public visit |

**Table 2. Summary of studies.**

we observed visitors informally and recorded their activity (video, electronic guidebook logs) while they used the electronic guidebooks. Also, in all studies, the designer (often in conjunction with the fieldworker or the conversation analyst) conducted semi-structured interviews with visitors after they used the electronic guidebooks. The three studies can be summarized as follows:

*Study 1 (October 2000).* Study 1 participants used Prototype 1, which offered a choice of delivery mode (text, audio through headphones, or audio through speakers). Study 1 included 14 individuals (seven pairs) from the Xerox PARC community. Participants visited two rooms on a day Filoli was closed to the general public.

*Study 2 (July 2001).* Study 2 participants used Prototype 2, which had eavesdropping. Study 2 included 12 individuals (six pairs) from the Xerox PARC community. Like Study 1, Study 2 was conducted on a day Filoli was closed to members of the general public. Participants visited three rooms on a day Filoli was closed to the general public.

*Study 3 (October 2001).* Study 3 participants used Prototype 2. Study 3 participants were 47 members of the general public (20 pairs, one group of three, and one group of four). These visitors were recruited on-site over four days when Filoli was open to members of the general public. Approximately half of the visitors approached agreed to participate, so there is of course a chance that self-selection is a factor. As in Study 2, participants visited three rooms. For a summary of the studies, see Table 2.

## CONVERSATION ANALYSIS

In making our case that CA can be useful as an integrated part of the design process, we will draw on specific aspects of the methodology and its outlook. In particular, we observe that CA provides findings of varying degrees of generality, and that these provide various kinds of design-relevant information. In this section, we briefly explain the methods and goals of CA, demonstrate how findings are produced (drawing on examples from our own studies, described in the last section), and provide a general discussion of how and when such findings might be relevant to designers.

### What is conversation analysis?

Conversation analysis, the most visible and influential form of ethnomethodological research, is concerned with describing the methods by which the members of a culture engage in social interaction. A key goal of conversation analysis is to examine social interaction to reveal organized practices or patterns of actions, under the fundamental assumption that interaction is structurally organized. Social actions include talk, gesture, and use of objects.

While ethnomethodology and conversation analysis share this concern for how actions are organized, the goal of CA is to describe both how sequences of action are organized and situated in a particular instance of activity, as well as to abstract features that *generalize* across a collection of similar instances.

A conversation analytic research program involves analyzing a collection of interactive encounters. The analysis is twofold. First, the analyst makes a moment-by-moment, turn-by-turn transcript of the actions in each encounter. Second, the analyst examines these encounters individually and then comparatively to reveal a practice's generalizable orderliness.

To make the discussion more specific, consider the procedure used in our own studies. In our project, the goal is to describe visitors' systematic practices as they use an electronic guidebook to tour a historic house with a companion. To identify these systematic practices, we examine in detail the data collected during selected visits. Specifically, for each visit, we create a video that includes the audio and video recordings of the visitors, as well as audio of the descriptions and video of the screens of each visitor's electronic guidebook. The resulting data are transcribed and analyzed. Both of these steps require careful, repeated viewing of each video.

### How does conversation analysis proceed from detailed to general findings?

While we have so far emphasized that CA is concerned with detail, we have also made assertions about the ability of CA to capture generalizations about practice. In this subsection, we present several excerpts taken from transcripts of our own studies. (Table 1 summarizes the notation used in the transcripts.) Using the excerpts, we demonstrate how instances of specific behaviors are used to identify a more general behavioral phenomenon. We then show how multiple phenomena can be generalized to make a higher-level point.

| **X:** <br> **X-PDA:** | Visitor X is speaking. **((comment or action))** <br> *Visitor X's guidebook is speaking.* |
| --- | --- |
| °soft°, °°whisper°° | Speech at reduced volume. |
| my [talk <br>    [your talk | Alignment of overlapping speech or actions. |
| my talk= <br>    =your talk | Latched (speech or actions with no interval between them). |

**Table 1. Summary of transcription notation.**

The excerpts and phenomena all relate to the ways in which the paired visitors incorporate the guidebook audio into their interaction. Note that these phenomena are practices that occur systematically, not hard rules. Therefore, there may be variations. However, these practices occur repeatedly with multiple pairs and represent a pattern of behavior.

*Phenomenon: speakers abort their turn when audio starts.* The analyst is attuned to certain kinds of practices because of her training. For example, the analyst noted many instances in which a human speaker stopped talking when a guidebook description started playing.

In Excerpt I, visitor A interrupts her introduction of a description when the audio begins playing.

***Excerpt I.***
```
1   A:       eh hm, this is a picture of uh mrs.=
2   A-PDA:                                      =This is a portrait
3            of Mrs. Roth painted by her friend Lloyd Sexton…
```

In this case, A stops completely; in Excerpt II, visitor W stops speaking when K's guidebook starts to play a description, resuming briefly to finish his sentence.

***Excerpt II.***
```
1   K:       oh, this is an option, ((gestures with PDA towards
2            window he is about to select))
3   W:       you see the paint peeling off of the ((points to wall))
4   K-PDA:   This side of the house [faces the extensive formal…
5   W:                              [plaster,
```

Aborted turns occurred whether the speaker or their companion played the description.

*Phenomenon: speakers place turns at possible completion points.* Visitors would sometimes speak briefly while the guidebook was playing. Visitors would commonly position these comments at points where the utterance-in-progress could be understood as complete (known as *transition relevance places*, or TRPs), e.g., the end of complete clauses or sentences. Sometimes these points were in fact the end of a description; if they were not, and the guidebook description continued, visitors would generally stop speaking.

Excerpt III shows an illustrative example, in which G starts to speak after the guidebook plays the word "Hawaii," thereby forming a possible complete utterance. From a collection of similar cases, we learn that visitors monitor the guidebook for possible TRPs and position their utterances in these positions, just as they do with human conversationalists.

***Excerpt III.***
```
1   R-PDA:   Mrs. Roth, the daughter of the founder of Matson
2            Navigation, collected these bowls from Hawaii,
3            [Fiji, and Tonga.
4   G:       [these bowls right here,
5            ((puts left arm on J's shoulder, turns to bowls))
```

*Phenomenon: speakers mark overt overlapping talk.* As mentioned above, visitors sometimes spoke while the guidebook audio was playing. In some cases, visitors explicitly marked their talk to show that they were talking in overlap with the guidebook, e.g., they whispered.

Excerpt IV exemplifies the use of whisper talk. Partway through the description playing on J's guidebook, L makes an observation in a whispered voice, and J responds in a similar manner. The whispering is not motivated by the volume (which is reasonably loud) or social considerations (since J and L are the only visitors present in the room). This excerpt is also an example of the use of TRPs mentioned in the previous point – L's utterance is targeted for the TRP after "bottomed out" and J delays her response to L's comment until another TRP occurs.

***Excerpt IV.***
```
1   J:       okay, so let's find this guy,
2   J-PDA:   Mr. William Bowers Bourn was the original owner
3            of Filoli. While he was attending Cambridge, the
4            Bourn family gold mine bottomed out. All
5            [visible ore was exhausted and engineers declared
6   L:       [°°lookit, he's a smoker,°° ((points to portrait))
7   J-PDA    the prospects grim.=
8   J:                                    =°oh, how funny,°=
9   J-PDA:                                                 =When
10           Bourn heard the news, he returned to California, ...
```

*Summary: speakers organize turns-at-talk with the guidebook audio in a manner similar to that in which they organize turns with a human speaker.* The preceding points all illustrate behaviors that humans are known to follow when in conversation with other humans. We have found a number of other behaviors that also demonstrate this point. Therefore, we see that visitors are generally orienting to the guidebook as though it is another conversational voice.

### *Why* might conversation analysis be useful?

As we have seen, CA findings are drawn from detailed examinations but ultimately comprise generalizable patterns of behavior. These patterns can inform the designer in several ways, depending on their degree of generality.

First, given a behavior of interest, CA can help isolate factors that contributed to the occurrence of that behavior. These factors can then be manipulated in the next design iteration. For example, if the designer wishes to increase instances of that behavior, they can try to make the factors occur more frequently; if the designer wishes to decrease instances of that behavior, they can try to make the factors occur less frequently.

Second, conversation analytic results can help the designer make predictions about how users will interact with new designs. The expectation is that findings of appropriate generality will continue to apply in the new situation.

Third, CA can set up a rigorous framework for comparing user behavior across design iterations. Put loosely, the analyst can isolate practices that can be compared even when different people, systems, and environments are being studied.

Fourth, a structured understanding of user behavior can confirm or deny that a design is achieving its desired objective. This is of particular interest when an objective can itself be expressed in terms of observable behavior, since a behavior can be verified even when subjective responses (from questionnaires, interviews, etc.) might be misleading.

The process of analyzing and viewing data can also lead to design intuitions. Certain user behaviors may inspire new designs. Further, analysis may bring into focus aspects of the design that have not explicitly been considered but nonetheless have dramatic effect on user behavior. Finally, an understanding of natural interaction patterns helps designers create artifacts that fit people's existing behaviors.

### *When* might conversation analysis be useful?
Given that CA is a specialized, resource-intensive methodology, it should be used only when appropriate. While this is by no means an exhaustive list, we have identified several situations in which we believe it is particularly useful.

First, we feel CA is especially helpful in the context of a broad design (or research) agenda. Because the goal of CA is to provide a rigorous analysis and generalizable results, rather than simply to get a specific technology to work, it yields insights that are helpful for designing an entire class of systems. Further, in a broad agenda, the high upfront costs of building an interaction framework and establishing working relationships between the analyst and the rest of the team can be amortized over the life of the project.

Second, we feel CA can yield some types of results that other methods can not. For example, certain behavior patterns are subtle and difficult to examine without CA, e.g., people do not self-report them accurately in interviews.

Third, we believe that since CA traditionally focuses on human participants, it is uniquely suited to the design of systems that enhance human-human interaction.

By contrast with the above situations, we observe that CA is not appropriate when similar effects can be achieved using a less resource-intensive method. For example, we feel that usability testing can generally be performed more efficiently without the use of CA.

## CASE STUDY: THE DESIGN OF AN AUDIO DELIVERY MECHANISM
Conversation analysis has impacted our design in a number of ways. In this section, we focus on the ways conversation analysis impacted the evolution of one aspect of the design, the audio delivery mechanism. We describe steps in the iterative design process, e.g., how the findings of a study influenced the design of the next prototype. While these phases are presented serially, many of them in fact overlapped and influenced each other. For example, Study 1 findings were in part developed concurrently with the design of Prototype 2.

In this section, we use callouts to draw relationships between the potential design benefits discussed in the CA section above and our experience with the design of the audio delivery mechanism. The callouts draw attention to particular examples of these benefits; however, many other instances that are not highlighted by callouts appear in this section as well.

### Prototype 1 design process
When we developed our first prototype, we simply knew that we were interested in designing an electronic guidebook that could enhance social interaction. To this end, we provided visitors different delivery modes to see how they affected social interaction.

### Study 1 findings
During Study 1, interviews revealed that visitors overwhelmingly preferred audio through speakers. Many said this preference was in part due to the fact that this delivery mode enhanced the social nature of their visit. Informal observations and interviews indicated that visitors would often discuss the content of the audio descriptions. Visitor response to the prototype overall was very positive.

CA helped us understand *how* speaker audio enhanced social interaction. CA revealed that the visitors oriented to the guidebook as though it was a human participant [27]. Visitors structured their conversations around the guidebook's audio, creating a place for it in their social interaction, e.g., visitors made a place for the guidebook to take turns in the conversation. This is an example of a general framework that the analyst needed to establish.

*Setting up the analytic framework*

Once this framework was established, the analyst was able to conclude that synchronized listening was a key factor in the social interaction, enabling visitors to integrate the guidebook in their existing conversational patterns. More specifically, CA demonstrated that visitors oriented to the guidebook descriptions as though they were stories, following discourse patterns that have previously been observed with human storytellers [20]. When a story ends, there is a place for listeners to respond to the story. We identified many such shared response moments in the Study 1 data. These shared response moments were a key factor in the social nature of the visits, and they were uniquely enabled by synchronized listening.

*Identifying key factors*

Regarding the structure of visitor activity, interviews indicated that visitors highly valued being able to select items independently and that visitors wanted to each have their own guidebook, because they wanted to control their own experience. At the same time, visitors said they enjoyed having a shared social experience. CA helped us understand the process by which visitors achieved both of these apparently conflicting goals, indicating that many visitors exhibited dis-engaging and re-engaging patterns, alternating between independent and shared activity. A prior study of re-engaging and dis-engaging talk in a peer group classroom setting [24] informed the analysis of these same phenomena in the studies of this project. This is precisely the value of using conversation analytic methods; while the analytic process may be time consuming, the findings of a particular data set can inform analyses in other studies.

**Prototype 2 design process**
We did not consider speaker audio a practical alternative for most public settings, since we expected it would be annoying when used by a large number of visitors. Therefore, our design challenge was to develop a prototype that would simulate the positive experience achieved with speaker audio but minimize the noise level in the room. For hardware, we settled on the use of single-earphone headsets and wireless networking to connect the guidebooks.

For the audio sharing model, we distilled several design constraints from our Study 1 findings and from existing knowledge of human-human interaction patterns. First, synchronized listening was necessary, as we learned from the conversation analysis. Second, the design had to support independent activity as well as synchronized listening. Third, the visitors should perform the minimum amount of work necessary to share, since extra work could interfere with the flow of their visit; this was indicated both by the conversation analysis of Study 1 data and by existing knowledge of human-human interaction.

We used these constraints to eliminate several classes of designs. For example, we dismissed asynchronous designs similar to email or instant messaging that would allow people to send each other audio clips. While these designs are fairly obvious and simple, and have in fact been suggested to us by a number of people, they fail to support the synchronized listening that the conversation analysis demonstrated was so important in Study 1. Further, most of the designs we considered in this category require users to take an explicit action each time they want to share an individual clip. The conversation analyst and the designer agreed that this explicit coordination would likely interfere with the flow of the visit. Accordingly, these asynchronous designs were ruled out. As another example, we ruled out systems in which users heard identical content at all times, since we wanted to support independent activity.

> *Predicting response to design changes*

Taking into account the design constraints, the team settled on the audio space model used in Prototype 2. This model supports synchronized listening, but gives individual control to each visitor. Further, it does not require coordination around each choice of object: visitors simply set the volume for eavesdropping and only change it when they desire, so they can listen to multiple clips together without taking any explicit action to share, beyond setting the volume in the first place.

**Study 2 findings**
In Study 2, we tested Prototype 2, which included the new eavesdropping model.

In the interviews, some pairs reported having a social experience, while others did not. Those who did report having a social experience generally described it in the same way that Prototype 1 speaker audio users had described their social experiences.

CA helped us understand which visitors had social experiences, the process by which those social experiences occurred, and the nature of those social experiences. Supplemented by interview data and informal observation, CA revealed that the most social pairs were generally those who participated in *mutual eavesdropping*, a phenomenon in which both participants in a pair use eavesdropping to listen to content from each other's guidebooks. CA allowed us to rigorously characterize the social interaction of mutual eavesdroppers, revealing strong patterns of visitor-visitor engagement that had not occurred in the previous study that used speaker audio [2]. For example, longer and more substantive response conversations occurred with mutual eavesdropping than with speaker audio. Further, mutually eavesdropping visitors tended to have more shared activity than those who used speaker audio.

> *Comparing practices across designs*

The conversation analysis led to the unexpected conclusion that mutual eavesdroppers were actually more cohered with technologically mediated, eavesdropped audio than with speaker audio. CA was uniquely qualified to help us make the comparison between the two prototypes; interviews did not provide insight into the structure of visitors' interactions, and since CA leads to generalizable results, we were able to compare the findings from Study 1 with those from Study 2.

CA indicated that non-mutual eavesdroppers were less cohered than mutual eavesdroppers, having less social interaction and engaging in more independent activity. The interviews did indicate that the non-mutual eavesdroppers found the eavesdropping useful for specific tasks. The team concluded that while Prototype 2 was successful for mutual eavesdroppers, it needed to be studied further in the case of non-mutual eavesdroppers.

> *Assessment of goal achievement*

**Study 3 design and analysis**
The interest in non-mutual eavesdroppers that resulted from Study 2 impacted the content of the interviews and informal observations of Study 3. The results of the interviews and informal observations in turn influenced the focus of the conversation analysis of Study 3 data, which is currently being conducted.

## INTEGRATING CONVERSATION ANALYSIS IN AN ITERATIVE DESIGN PROCESS

Development teams that include qualitative analysts often employ a consulting relationship. In this model, the qualitative analysts do their work separately from the rest of the team; findings are communicated as work is completed. While some instances of this model involve regular meetings between the developers and the qualitative analysts, e.g., for qualitative analysts to present their results to the developers or to answer questions, the analysts are generally not fully integrated in the design process.

However, we believe we have achieved excellent results by fully including the conversation analyst in the team, and therefore, fully integrating CA in the iterative design process. In this section, we discuss some of the strategies we have developed for achieving this integration.

### Integrating the conversation analyst

We use several organizational strategies to fully integrate the conversation analyst in the design team.

*Pairing the analyst with the designer.* Close collaboration between the conversation analyst and the designer is in large part the key to our successful integration of the analyst in the team. This collaboration is characterized by extremely frequent informal communication and one-on-one meetings in which the conversation analyst and the designer share findings and work together on both analysis and design. We discuss aspects of this interaction further below.

*Educating the designer.* Before the project began, the designer was unfamiliar with conversation analytic methods. To familiarize herself with the methods, the designer apprenticed herself to the conversation analyst, and actively participated in the analysis of data from Study 1. The designer found it helpful to set herself specific exercises that contributed to the analysis, e.g., to look for all instances of particular movement patterns. In Study 2 analysis, the designer watched the videos in their entirety a small number of times and looked at specific excerpts identified by the conversation analyst. This decreased level of involvement was largely due to time constraints, but the team was pleased to discover that it was very productive. We believe it worked because the designer had participated in the previous analysis and was therefore familiar with both the conversation analytic method and the interaction framework developed during the analysis of Study 1 data.

Note that the designer does not feel that it is generally feasible for a designer to simply do conversation analysis on their own. First, the skills required take many years to acquire, and most designers do not already have this extensive training; while the designer now has some familiarity with the techniques, she certainly does not feel qualified to conduct such inquiry on her own. Second, the technique is very time-consuming.

*Including the analyst in formal and informal interactions with the team.* Formal and informal interaction between the conversation analyst and the rest of the team is frequent. The conversation analyst participates in most project meetings, particularly those related to interaction design. Because the designer is deeply familiar with the conversation analyst's work, she can help advocate points or serve as an interpreter for other members of the team.

*Legitimizing and communicating the role of conversation analysis.* A further advantage of the complete integration of the conversation analyst is the team is that she does not feel that she has to explicitly justify the value of her work, since the work is an outcome of a group process. The conversation analyst contrasts this with a consulting model in which her contribution might frequently be questioned or misunderstood.

*Providing technical assistance and infrastructure.* Applying CA in a design setting may require equipment and skills that are not typically used or taught in academic environments. Teams should expect to provide significant assistance in terms of time and resources.

In academia, CA is typically conducted using single-track audiovisual recordings. In the course of analysis, these recordings are replayed many times. As a result, analysts are likely to be facile with consumer digital video recording equipment, digital video capture, and media playback software. Video editing skills are not usually required.

By contrast, our studies required the analyst to understand not only the participants' actions, but the technological context of their actions as well. Specifically, the analyst needed to know what the visitors were seeing and hearing from the guidebook at all times. To enable the analyst to use media playback tools familiar to her,[1] the team assembled a single composite video for each visit (Figure 2). Each frame of the composite video contained three active video tracks, two of the device displays and one of the visitors (cutting between the 2-4 tracks of the visitors available for each visit). In addition, the composite video contained four active audio tracks: two of the device audio and two from the wireless microphones carried by each visitor. Fairly precise track synchronization was necessary because temporal alignment of events is a critical part of CA; for example, it matters whether a visitor's comment is delivered during or after the playback of a guidebook description.

---

[1] Another alternative would have been to write software that provided an integrated, easy-to-use video analysis environment (see, e.g., [12,16]). This was beyond the resources of our project.

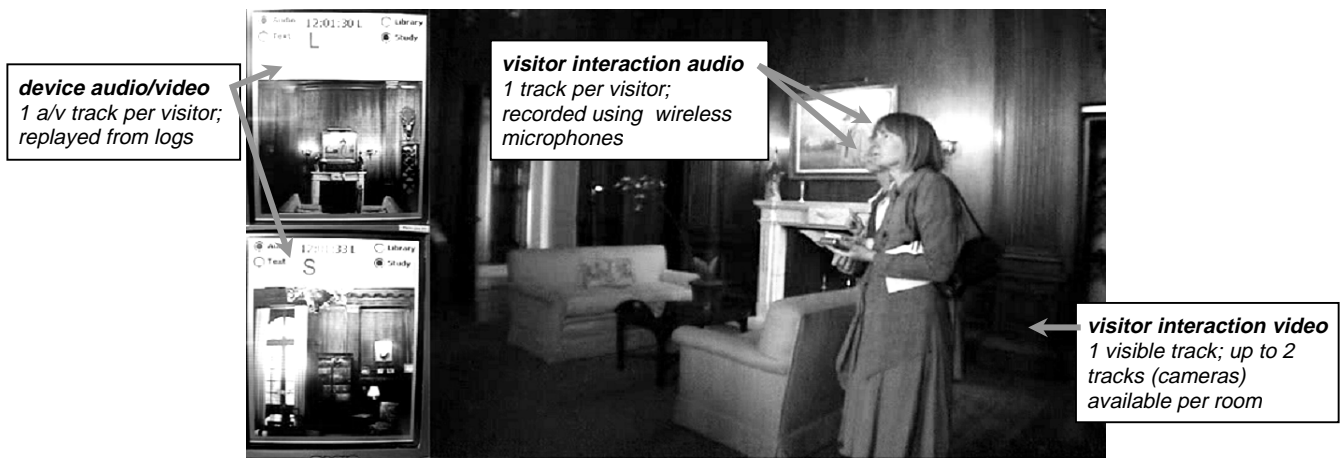

**Figure 2. Putting together the composite video for the analyst.**

Producing this kind of video requires planning. Resources must be allocated for audiovisual recording and editing equipment. Prototypes must be instrumented to capture necessary information. Finally, the team will likely spend time assisting with the design and execution of the study.

**Managing the analysis process**
As we have mentioned previously, conversation analysis is a very time-consuming and detailed method. Further, given a data set such as ours that is extremely rich, undirected conversation analysts could spend literally years studying myriad aspects of the data. However, the conversation analyst can also be selective, choosing only specific avenues of exploration. A selective analysis can provide relevant results on a more reasonable time scale without compromising the intellectual integrity of the method. In this subsection, we discuss some of the issues about deciding what work to do and who should do it.

*Maintaining analytic breadth.* We recognize that the conversation analyst must do some work that may not be of immediate relevance to the design. They must pay an up-front cost to establish an analytic framework for the type of interaction that is occurring in a particular domain. This framework is the foundation for later findings that are immediately relevant to the design; the need for this framework must temper attempts to focus early in the process [17].

A related point is somewhat specific to the research environment. As recognized by Sommerville [23] (among others), members of an interdisciplinary team must be able to pursue their own research agendas. This means that sometimes the conversation analyst does work that does not have clear applicability to design issues but makes a valuable academic contribution.

*Maintaining analytic focus.* While the conversation analyst needs latitude to do different types of work, on our project she is highly motivated to study issues that will help with the design process. The challenge is not to find problems that are interesting to her (since there are many), but rather to help her identify research questions that will inform design. The designer facilitates this process in two ways: (1) the designer listens to all of the conversation analyst's preliminary findings and helps her decide which issues are most worth pursuing and (2) the designer asks questions that are intended to prompt the conversation analyst to consider relevant issues.

In addition to focusing the analysis on particular issues, it is also useful to focus the analysis on particular subsets of the data, particularly since creating the video composites is extremely time-consuming, and because we find it useful to interview more pairs than we can realistically analyze using conversation analytic methods (we can get different types of information from interviews, but interviews typically require more participants). This is one reason it is very helpful to have the designer make in situ observations and conduct interviews – the designer can help identify the most promising pairs to study. While we do not do a full analysis of all visitors, we find it useful to study in-depth the pairs that we do select; we find it particularly informative to study the entire visit for selected pairs, because in this way we can see each action in the context of the entire visit, e.g., when one visitors plays a description, we know whether the other visitor has heard it already.

*Managing participation.* Our approach does not seem to require more than two people (one conversation analyst and one designer), although since conversation analysis is so intensive we think it might be productive to add another conversation analyst. The conversation analyst and the designer do find it very productive to review the transcripts and video with other conversation analysts at Xerox PARC (cf. [8, p.194]); group analysis meetings, or "data sessions," are a standard technique for conversation analysts.

**Bridging interdisciplinary gaps**
In most interdisciplinary interactions, collaborators must work to communicate with each other and to understand each other's methods. In this section, we discuss some of the particular challenges we have faced and ways we have addressed them.

*Addressing mismatches in vocabulary.* Many issues, including differences in methodology and vocabulary, can make it difficult for social scientists and engineers to communicate (see, e.g., [19,23]). Conversation analysts have a vocabulary that can be especially misleading because conversation analysts assign very specific meanings to common words, e.g., "repair," "sequence," and "phenomenon" are all precisely defined and have particular connotations. Team members must be careful not to make assumptions about meanings of terms and they should ask for definitions as new terms arise.

*Understanding what CA findings might provide.* Initially, the designer did not know what types of questions to ask the conversation analyst. Over time, as the designer has become more familiar with conversation analysis and gained more experience with the types of results that are helpful for design, she has learned to ask better questions. Specifically, she has learned to ask questions about process and causality.

*Understanding what the design process requires.* At the beginning, the conversation analyst did not have a sense of what types of results would be interesting. She has gained a better sense of this. Further, she has realized that causality is an important factor for design. The conversation analyst reports that this type of thinking is new to her – she generally starts by saying, "I saw this phenomenon" and was initially surprised when members of the team would ask her what caused it. This process is now more familiar and comfortable for her.

**Communicating within the group**

As we have discussed above, frequent communication among all members of the group is important. However, it is obviously infeasible for all members of the group to discuss all possible findings and ideas. In this section we discuss the communication patterns that have evolved in our team.

*Sharing of nascent ideas during analysis.* The conversation analyst generally shares all findings (even preliminary ones) with the designer, who takes notes on findings that she thinks may be relevant to design. Further, the designer shares most design suggestions with the conversation analyst. This sharing occurs during frequent scheduled and unscheduled meetings and phone calls; many days the designer and conversation analyst will talk briefly just to give each other progress reports. The thoughts shared are often tentative, e.g., "today, I was thinking maybe it would be a good idea to try a design that did the following…" or "I was just looking at the data, and I'm not sure, but it seems as though the visitors might be doing the following…" Further, the designer and the conversation analyst share most intermediate products, e.g., transcripts, diagrams of episodes, notes, etc.

The informal and frequent nature of the collaboration has many advantages. Because we are in constant contact, we are checking each other's conclusions and decisions every step of the way. The conversation analyst constantly provides input on the design as it evolves, and the designer constantly interacts with the conversation analyst about the most productive lines of inquiry. In less frequent interactions, the conversation analyst has to do more "second guessing" about what will be relevant to the rest of the group. When the designer is not involved, the entire filtering burden falls on the conversation analyst, which means the conversation analyst runs the risk of focusing on less relevant issues and never bringing more relevant issues to the attention of the designer and the rest of the team.

After the designer and the conversation analyst identify interesting findings, they bring them to the rest of the team. The team may ask follow-up questions of the conversation analyst. New designs are proposed by the conversation analyst and the designer, as well as by other members of the team, and the entire team considers whether they expect these designs to be successful. The team members bring their experience to bear in this stage, e.g., the conversation analyst uses her understanding of human-human interaction as well as her understanding of the data from the previous studies, while the designer uses her understanding of human-computer interaction as well as her understanding of the data from the previous studies.

*Selecting appropriate presentation methods for CA findings.* We do not generally find it is necessary for the other members of the team to view the video data or the transcripts. However, the designer and the conversation analyst often use storytelling [9] during team meetings, e.g., they describe specific episodes and the team discusses how the design could accommodate the behaviors that occurred. Further, the conversation analyst does on occasion share intermediate artifacts with the whole team. For example, annotated transcript excerpts and abstract "flowchart" diagrams of a visit have been helpful in explaining specific interactions.

**Relating complementary study/analysis methods to conversation analysis**

We use complementary methods for a number of tasks on our project.

*Accomplishing tasks not appropriate to CA.* We use complementary methods to accomplish tasks that are not well-suited to conversation analysis. Often these tasks could potentially be achieved by CA, but can be performed much more efficiently with other methods. For example, to improve the usability of the system, we do task-based usability testing. We can perform this usability testing much more efficiently than we could conduct a conversation analytic study, even though CA can be used to achieve similar results [8]. (Also note that we perform this usability testing before conducting a conversation analytic study of our prototypes, since we believe usability problems are distracting and influence the character of the visits.)

As another example, certain kinds of participant responses, such as participants' subjective impressions of how much they enjoyed their visit, can not be observed, and therefore, are most appropriately acquired through interviewing.

*Informing CA.* We use complementary methods to inform the conversation analysis. Information from interviews, in situ observations, and log visualizations help us identify questions for analysis, and they also help us identify visitors who will be interesting to study (i.e., which recordings to analyze).

*Informing interview questions.* Just as the interviews inform the conversation analysis, the conversation analysis identifies interesting issues to be pursued in interviews.

**RELATED WORK**

Designers have been working with qualitative social scientists for over a decade. Reports on this work often appear in the (overlapping) areas of HCI, computer-supported cooperative work (CSCW), and computer-mediated communication (CMC).

The application of CA to system design has been the major subject of several books [15,25,28] and a number of papers. There are two main ways that CA has been applied. First, designers have attempted to use CA as a source of principles (commonly referred to as conversational "rules" or "protocols") for making systems more "conversational" in their interactions. For example, CA has been used for modeling conversation in CMC systems [1,5,22] and interactive systems simulating human conversational patterns [7,10]. Other researchers have attempted to apply such "rules," essentially by analogy, to interactive systems of other kinds [11,18,26]. Second, CA has been used as an experimental "evaluation" method for screen-based interfaces [8,18], speech interfaces [28] and CMC systems such as collaborative virtual environments (CVEs) [6,13]. That is, users are recorded as they attempt to use a system, and their interactions are examined using conversation analytic techniques.

In spite of this activity, there are no detailed, "how to" discussions of the involvement of conversation analysts in the iterative design process. There have been a variety of reports of experiences involving ethnographers (e.g., [4,14,23]); many issues raised in these reports do generalize, but (as we have discussed) application of CA raises its own particular issues. In one report [8], a designer applied CA-derived methods to address usability issues during iterative design. However, the designer/analyst was entirely concerned with using detailed examination to understand the *particulars* of specific problematic human-computer interactions – "to interpret what was really going on in the interactions" [8, p.190] – as opposed to making generalized findings about the organization of interactions, which is the goal of CA. Perhaps more to the point, a scenario involving a single person who plays the role of both designer and analyst reveals little about the mechanics of bringing a professional social scientist into a design team.

**CONCLUSIONS**

We have presented a case study in integrating a conversation analyst in an iterative design process. We have discussed potential benefits of CA for the design process and illustrated these benefits in an example from our own experience.

We have discussed the strategies we have used to integrate the analyst in the design team, manage the process of analysis, bridge interdisciplinary chasms, communicate within the group, and relate complementary methods to conversation analysis.

Using these strategies, we have found conversation analysis to be highly beneficial. As a result, conversation analysis has become central to our design process, and we enthusiastically plan to continue using it in the future.

**ACKNOWLEDGMENTS**

We are deeply indebted to Tom Rogers and Anne Taylor of Filoli for their generous assistance with this project. We thank Jim Thornton and Amy Hurst for their participation in the design, development, and evaluation process. We are grateful to Bob Moore, Morgan Price, Peter Putz, Terkel Skaarup, Ian Smith, Erik Vinkhuyzen, and Marilyn Whalen for their helpful insights.